# Are the *calorimetric* and *elastic* Debye temperatures of glasses really different?

M. A. Ramos

*Laboratorio de Bajas Temperaturas, Departamento de Física de la Materia Condensada,*

*C-III, Instituto de Ciencia de Materiales "Nicolás Cabrera",*

*Universidad Autónoma de Madrid, E-28049 Madrid, Spain.*

ABSTRACT

Below 1 K, the specific heat $C_p$ of glasses depends approximately linearly on temperature $T$, in contrast with the cubic dependence observed in crystals, and which is well understood in terms of the Debye theory. That linear contribution has been ascribed to the existence of two-level systems as postulated by the Tunnelling Model. Therefore, a least-squares linear fit $C_p = C_1 T + C_3 T^3$ has been traditionally used to determine the specific-heat coefficients, though systematically providing *calorimetric* cubic coefficients exceeding the *elastic* coefficients obtained from sound-velocity measurements, that is $C_3 > C_{Debye}$. Nevertheless, $C_p$ still deviates from the expected $C_{Debye}(T) \propto T^3$ dependence above 1 K, presenting a broad maximum in $C_p / T^3$ which originates from the so-called *boson peak*, a maximum in the vibrational density of states $g(\nu)/\nu^2$ at frequencies $\nu \sim 1$ THz. In this work, it is shown that the apparent contradiction between *calorimetric* and *elastic* Debye temperatures long observed in glasses is due to the neglect of the low-energy tail of the boson peak (which contribute as $C_p \propto T^5$, following the Soft-Potential Model). If one hence makes a quadratic fit $C_p = C_1 T + C_3 T^3 + C_5 T^5$ in the physically-meaningful temperature range, an agreement $C_3 \approx C_{Debye}$ is found within experimental error for several studied glasses.

It is well known that glasses exhibit physical properties very different from those of crystalline solids, especially at low frequencies or low temperatures (Phillips 1981, Esquinazi 1998). Below 1 K, the specific heat $C_p$ depends approximately linearly on temperature $T$ (Zeller and Pohl 1971, Stephens 1976, Phillips 1981). This is in clear contrast with the cubic dependence observed in crystals which is well understood in terms of the Debye theory, which predicts in the limit of very low temperatures a specific heat given by $C_p = C_{Debye}\, T^3$, with

$$C_{Debye} = \frac{2\pi^2}{5}\left(\frac{k_B^4}{\hbar^3 \rho v_D^3}\right) = 234 \frac{nk_B}{\rho \Theta_D^3} \tag{1}$$

where $\Theta_D$ is the Debye temperature, $n$ is the number density of atoms, $\rho$ is the mass density, and $v_D$ is the average Debye sound velocity defined by

$$\frac{3}{v_D^3} = \frac{1}{v_L^3} + \frac{1}{v_T^3} \quad, \tag{2}$$

$v_L$ and $v_T$ being the longitudinal and transverse sound velocities, respectively.

The above-mentioned quasi-linear contribution to the specific heat at very low temperatures, as well as several other *universal* low-temperature properties of glasses or amorphous solids (Phillips 1981), have long been successfully accounted for by the Tunnelling Model (Anderson *et al.* 1972, Phillips 1972), which postulated the ubiquitous existence of atoms or small groups of atoms in amorphous solids which can tunnel between two configurations of very similar energy (two-level systems, TLS).

Nevertheless, above 1 K the specific heat of glasses still deviates strongly from the expected $C_{Debye} \propto T^3$ dependence, exhibiting a broad maximum in $C_p/T^3$. It is now accepted that this universal feature is related to a difference or excess in the vibrational density of states $g(\nu)$ over the crystalline Debye behaviour, leading to a maximum in $g(\nu)/\nu^2$ at frequencies $\nu \sim 1$ THz (Buchenau *et al.* 1986) which has become known as the *boson peak*, a dominant feature in the vibrational spectra of glasses. The very nature of the boson peak remains one the major debated and unsolved problems of condensed matter physics. Since we are not going to address directly this issue, suffice it to say that one may distinguish two basic kinds of approaches. On the one hand, several authors assume the coexistence at low enough frequencies of Debye-like acoustic phonons with *excess* vibrational excitations responsible for the boson peak (and with TLSs responsible for the properties at the lowest temperatures), as the Soft-Potential Model (SPM) does (Karpov *et al.* 1983, Il'in *et al.* 1987; for reviews, see Parshin 1994, Ramos and Buchenau 1998) or as Engberg *et al.* (1998), who found from inelastic neutron scattering experiments in $B_2O_3$ that the boson-peak vibrational motions could be decomposed in terms of in-phase and random-phase components: the latter, optic-like, would correspond to the density of states observed in Raman scattering, and the former would be the sound waves, whose density of states extrapolates approximately to the Debye value at low enough energies. Furthermore, the boson peak has also been considered to arise from the high-$Q$ limit of the transverse branch of acoustic phonons (Pilla *et al.* 2002). On the other hand, other authors consider that disorder in glasses spoils the picture of extended plane waves and hence makes the Debye theory unapplicable in the whole frequency range of the boson peak feature, which would originate from the softening of acoustic-like excitations by static disorder (Elliott 1992), marking somehow the crossover from propagating sound waves to localized vibrations, or from hybridization between acoustic and optic modes, etc.

However, since in the 1970s many experiments on thermal and acoustic properties at low temperatures were performed, and the rapid success of the Tunnelling Model (Anderson *et al.* 1972, Phillips 1972) to account for the universal properties of glasses below 1 K was in clear contrast with the poor understanding of the behaviour at higher temperatures or energies, the role of the *boson peak* as a basic feature of low-energy dynamics of glasses was rather put apart in those low-temperature research studies. Therefore, a least-squares linear fit $C_p = C_1 T + C_3 T^3$ in a $C_p/T$ vs $T^2$ representation for data (see figure 1a) below around 1 K has been traditionally used to determine the specific-heat coefficients and hence the density of tunnelling states and the Debye-like cubic contribution. Without exception, this direct method has provided (see the table on p. 37 of the book edited by Phillips (1981), largely taken from Stephens (1976)) *calorimetric* cubic coefficients clearly exceeding the *elastic* coefficients obtained from the Debye theory in those cases where sound-velocity measurements are available; that is, the statement $C_3 > C_{Debye}$ pervaded the scientific literature in the field for decades. As a consequence, the applicability of the Debye theory and the idea of well defined, extended sound waves within the amorphous lattice in the relevant frequency range (say, ~100 GHz) have naturally been questioned.

Nowadays, one of the approaches which most often used to understand the also rich and universal glassy phenomenology of thermal properties above 1 K, or vibrational dynamics above 100 GHz, is the aforementioned SPM. It can be considered as an extension of the Tunnelling Model, though the SPM is not so unanimously accepted as that model. In brief, the SPM (Karpov *et al.* 1983, Il'in *et al.* 1987) postulates the coexistence in glasses of acoustic phonons (crystalline-like, extended lattice vibrations) with quasilocal low-frequency (*soft*) modes. In the SPM, the potential of these soft modes is assumed to have a general form

$$V(x) = W\left(D_1 x + D_2 x^2 + x^4\right) \tag{3}$$

where $x$ is any generalized spatial coordinate, and the energy $W$, the stabilizing fourth-order term assumed to be the same for all atomic potentials, constitutes the basic parameter of the model. Each mode has its own first-order (asymmetry $D_1$) and second-order (restoring force $D_2$) terms, which can be either positive or negative, hence giving rise to a distribution of double-well potentials (TLS) and more or less harmonic single-well potentials (soft vibrations). The Tunnelling Model hence results included as a subpart of the SPM. The parameter $W$ marks the crossover from the TLS-dominated region at the lowest temperatures to the soft-modes region above it for the different physical properties under study. Very good agreement has been found between the SPM predictions and the experimental data (for reviews, see Parshin 1994, Ramos and Buchenau 1998). Similarly to the Tunnelling Model, a random distribution of potentials is assumed in the SPM: $P(D_1, D_2) = P_s$. As a result, the double-well potentials (TLS) contribute a quasi-linear in temperature specific heat

$$C_{p,TLS}(T) = \frac{\pi^2}{6}\left(\frac{1}{9}\right)^{1/3} P_s k_B \left(\frac{k_B T}{W}\right) \ln^{1/3}\left[\frac{t_{\exp}}{\tau_{min}(T)}\right] \tag{4}$$

where $t_{\exp}$ is the experimental time and $\tau_{min}$ the relaxation time for symmetric double-well potentials of a given energy, as in the Tunnelling Model. On the other hand, the density of vibrational excitations from single-well potentials (*soft modes*) is found to be proportional to $v^4$ and hence contributes to the specific heat, in the quasiharmonic approximation,

$$C_{p,sm}(T) = \frac{2\pi^6}{21} P_s k_B \left(\frac{k_B T}{W}\right)^5 \tag{5}$$

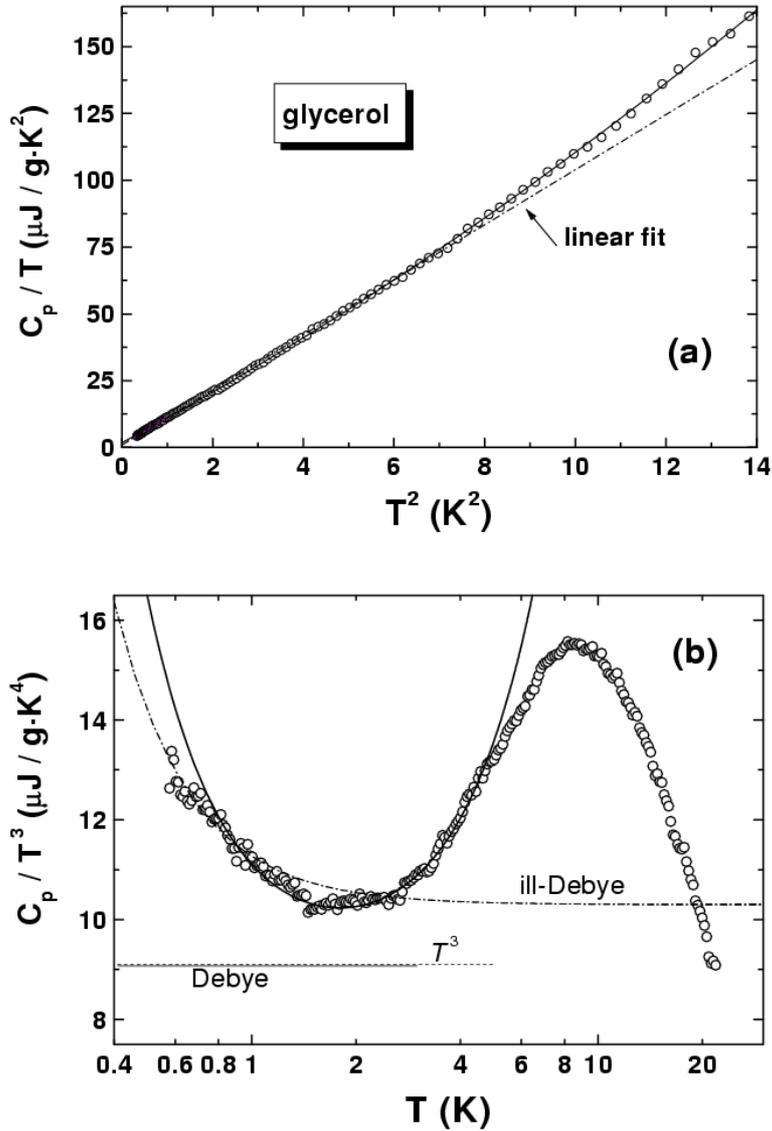

**Figure 1**. Specific-heat data of glycerol (Talón *et al.* 2002), together with a traditional linear fit (dashed lines) and the proposed SPM quadratic fit (solid lines). (a) $C_p/T$ vs $T^2$ plot, where both fits are carried out. (b) $C_p/T^3$ vs $T$ representation. The cubic term of the quadratic fit (dotted line) is compared with the predicted Debye value (solid line).

Both regions in the convenient $C_p/T^3$ representation are clearly separated by the crossover $T_{\min}$, the temperature of the minimum in $C_p/T^3$, that is directly related to $W$ through $W \approx$ (1.8—2) $k_B T_{\min}$. Of course, at sufficiently high energies or temperatures, the assumption of independent quasiharmonic soft modes, coexisting with independent sound waves, and hence equation (5), will cease to be valid and their mutual interactions will somehow produce (Gurevich *et al.* 2003) the maximum in $C_p/T^3$ (the boson peak).

It will now be shown that it is precisely the neglect of the low-energy tail of the boson peak (which contributes as $C_{p,sm} \propto T^5$, equation (5) following the SPM, or also from more general considerations, see Gurevich *et al.* 2003) that causes the *apparent* discrepancy between the measured (*calorimetric*) and the calculated (*elastic*) Debye coefficients of the specific heat. Although this soft-mode contribution might seem not very important in $C_p/T$ vs $T^2$ representations (see figure 1a) below 1 K, its neglect strongly biases specific-heat data fits towards lower temperatures, as is better observed in a $C_p/T^3$ vs $T$ plot (see figure 1b). Moreover, $C_1$ and especially $C_3$ depend on the chosen range for the linear fit. In order to solve these problems, we have proposed (Ramos *et al.* 2003) a systematic method to analyse low-temperature specific-heat measurements, which basically consists of carrying out a quadratic polynomial fit to $C_p/T$ versus $T^2$:

$$C_p(T) = C_{TLS}T + C_D T^3 + C_{sm} T^5 . \qquad (6)$$

The question however remains how to decide the temperature range to fit the data, with a sound physical meaning. To be consistent with the assumptions made, the upper temperature limit should be placed where the distribution of independent soft modes begins to fail and correspondingly the specific heat starts deviating from the $T^5$ law, somewhere between $T_{min}$ and the boson peak maximum $T_{max}$. Based upon some SPM numerical calculations, we first proposed (Ramos *et al.* 2003) to fix $(3/2)T_{min}$ as such upper limit. On the other hand, Gurevich *et al.* (2003) have recently shown that a vibrational instability of any unstructured spectrum of weakly interacting quasilocal harmonic modes naturally produce the appearance of the boson peak feature, with an excess of the density of states over the Debye value again proportional to $\nu^4$ up to approximately one half of the position of the boson peak. As an averaged compromise between both approaches, we suggest here performing the fit to equation (6) in the temperature range $0 < T < (3T_{min} + T_{max})/4$. Figure 1 shows this applied to glycerol data (Talón *et al.* 2002) compared to the traditional linear fit at lower temperatures. As can be observed, the cubic coefficient $C_D$ so determined agrees very well, within experimental error, with that predicted by the Debye theory from elastic data, what does not occur with the "ill-determined" Debye coefficient $C_3$ obtained from the traditional linear fit.

Figure 2 shows this SPM quadratic fit carried out on a system of sodium-oxide borate glasses $(B_2O_3)_{100-x}(Na_2O)_x$ earlier measured by our group (Piñango *et al.* 1990, Ramos *et al.*

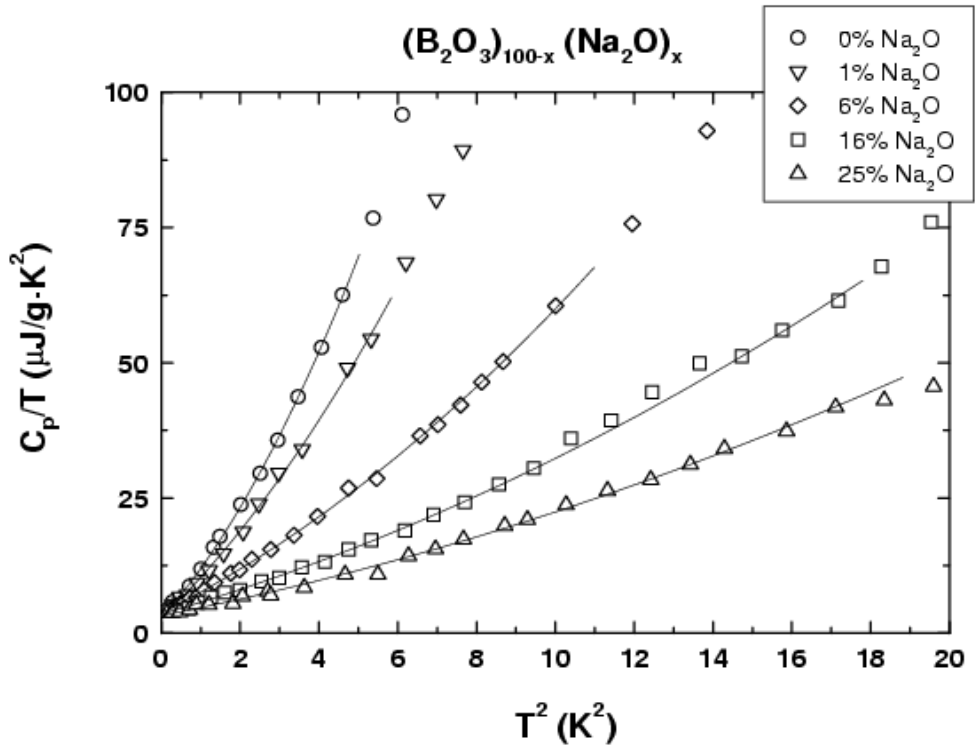

**Figure 2**. Specific-heat data of $(B_2O_3)_{100-x}(Na_2O)_x$ glasses (Piñango *et al.* 1990, Ramos *et al.* 1990), together with SPM quadratic fits (solid lines).

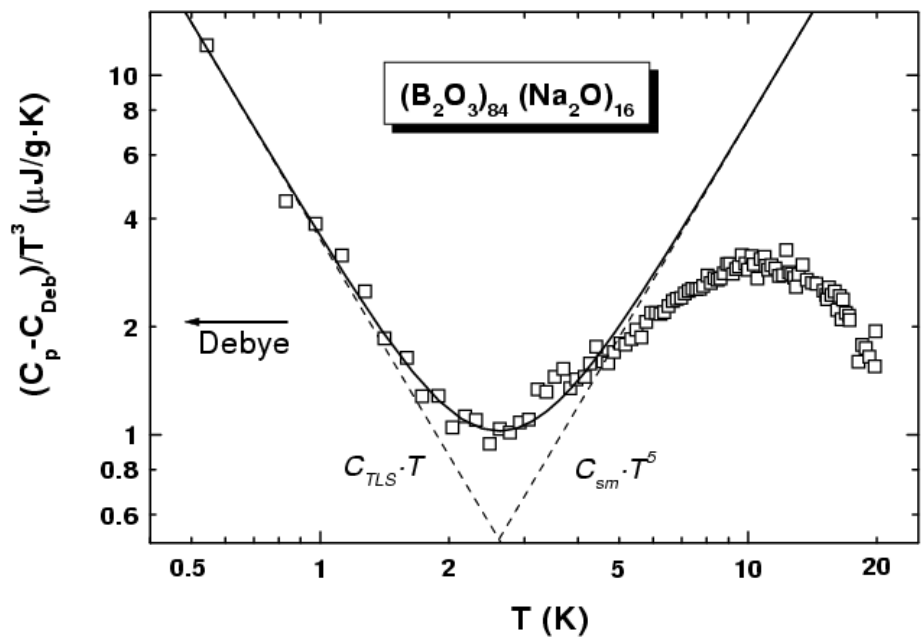

**Figure 3**. Specific-heat data of the $(B_2O_3)_{84}(Na_2O)_{16}$ glass (Piñango *et al.* 1990), in a log-log $C_p/T^3$ vs $T$ representation after subtraction of the Debye term. The solid line corresponds to the fitting curve obtained in figure 2, and dashed lines are the separate contributions from TLS ($\propto T$) and soft modes ($\propto T^5$).

1990), and where the traditional linear fit at lower temperatures led us to claim that the cubic contributions to the specific heat systematically exceeded the predicted Debye values determined from elastic measurements after Krause and Kurkjian (1978), that is $C_3 > C_{Debye}$. In figure 3, the experimental data and the fit for $x=16$ are shown in a $C_p/T^3$ vs $T$ log-log plot, in this case after subtraction of the Debye contribution to emphasize the crossover from TLS to soft-mode regimes, as well as their combined contribution in the relevant temperature range around $T_{min}$.

| | **ELASTIC DATA** | | | | | | ← SPM quadratic fit → | | | |
|---|---|---|---|---|---|---|---|---|---|---|
| MATERIAL | $\rho$ (g/cm$^3$) | $V_l$ (km/s) | $V_t$ (km/s) | $C_{Debye}$ (µJ/g·K$^4$) | $T_{min}$ (K) | $T_{max}$ (K) | $C_{TLS}$ (µJ/g·K$^2$) | $C_D$ (µJ/g·K$^4$) | $C_{sm}$ (µJ/g·K$^6$) | ($C_D/C_{Debye}$) |
| SiO$_2$ | 2.20 | 5.8 | 3.8 | **0.77** | 2.1 | 10 | 1.34 | **0.77** | 0.091 | **1.00** |
|  |  | 5.8 | 3.7 | **0.83** |  |  |  |  |  | **0.93** |
| Se | 4.3 | 2.0 | 1.05 | **17.6** | 0.6 | 3.1 | 0.58 | **17.7** | 3.98 | **1.01** |
| PMMA | 1.18 | 3.15 | 1.57 | **19.0** | 1.5 | 3.6 | 5.71 | **19.0** | 1.69 | **1.00** |
| PS | 1.06 | 2.8 | 1.34 | **33.7** | 1 | 3 | 5.69 | **34.6** | 4.59 | **1.03** |
| CaK(NO$_3$)$_3$ | 2.1 | 3.456 | 1.749 | **7.73** | 2.3 | 6 | 5.39 | **7.37** | 0.171 | **0.95** |
| B$_2$O$_3$ (quenched) | 1.81 | 3.39 | 1.87 | **7.47** | 1.1 | 5.2 | 2.76 | **8.15** | 1.04 | **1.09** |
| B$_2$O$_3$ (annealed) | 1.85 | 3.62 | 2.00 | **5.97** | 1.1 | 5.5 | 1.59 | **4.48** | 1.03 | **0.75** |
| (B$_2$O$_3$)$_{99}$(Na$_2$O)$_1$ | 1.842 | 3.71 | 2.04 | **5.65** | 1.4 | 5.5 | 2.83 | **6.95** | 0.547 | **1.23** |
| (B$_2$O$_3$)$_{94}$(Na$_2$O)$_6$ | 1.960 | 4.36 | 2.27 | **3.81** | 1.9 | 7.5 | 3.75 | **3.65** | 0.197 | **0.96** |
| (B$_2$O$_3$)$_{84}$Na$_2$O)$_{16}$ | 2.122 | 5.08 | 2.79 | **1.92** | 2.5 | 10 | 3.50 | **2.13** | 0.075 | **1.11** |
| (B$_2$O$_3$)$_{75}$(Na$_2$O)$_{25}$ | 2.288 | 5.56 | 3.10 | **1.30** | 2.6 | 11 | 3.24 | **1.44** | 0.048 | **1.11** |
| Glycerol | 1.42 | 3.71 | 1.89 | **9.07** | 2.0 | 8.7 | 1.83 | **9.10** | 0.176 | **1.00** |

**Table 1**. Comparison of elastic and specific-heat data for several glasses, including the SPM quadratic fit of equation (6). The elastic data are as follows: SiO$_2$, Se, PMMA, PS, and CaK(NO$_3$)$_3$, taken from the review article by Pohl *et al.* (2002); (B$_2$O$_3$)$_{100-x}$(Na$_2$O)$_x$, from Krause and Kurkjian (1978); glycerol, after Ramos *et al.* (2003). The specific-heat data are as follows: SiO$_2$, from Zeller and Pohl (1971) and Buchenau *et al.* (1986); Se, from Brand and Löhneysen (1991); PMMA and PS, from Choy *et al.* (1970) and Stephens *et al.* (1972); CaK(NO$_3$)$_3$, from Sokolov *et al.* (1997); (B$_2$O$_3$)$_{100-x}$(Na$_2$O)$_x$, from Piñango *et al.* (1990) and Ramos *et al.* (1990); glycerol, from Talón *et al.* (2002).

Last but not least, one should check that the analysis proposed is consistent and meaningful, and that it not merely obtains a better fit in a wider range by introducing one more parameter. Firstly, and most importantly, we show that the unveiled agreement between the Debye coefficients obtained from elastic data and from the fits to calorimetric data is absolutely general, as can be seen in table 1. In this table, apart from the glasses discussed above, specific-heat data are included for five other typical glasses (which already appeared in the aforementioned classical tables, and exhibited cubic coefficients with systematic and large excesses over the Debye values). The SPM quadratic fit of equation (6) has been applied for all studied glasses in the temperature range $0 < T < (3T_{min} + T_{max})/4$, and compared with most recent available elastic data (see table heading for references). The last column of table 1, that is $C_D/C_{Debye} \approx 1.0$, provides evidence that the starting hypothesis of coexistence of additional, quasilocal modes with usual Debye acoustic phonons is well fulfilled within experimental error. The least favourable cases of an annealed $B_2O_3$ glass and other slightly doped with 1% $Na_2O$ can be traced back to the greater uncertainty in the elastic data, since the sound velocity of this glass is known to exhibit a very strong dependence on thermal treatments and water content (Ramos *et al.* 1997). Let us note that the $C_3/C_{Debye}$ ratios obtained by using the traditional linear fit (see table on p. 37 of the book edited by Phillips (1981)) for $SiO_2$, Se, PMMA, PS, and $CaK(NO_3)_3$ are 2.25, 1.17, 1.65, 1,74 and 1.86, respectively.

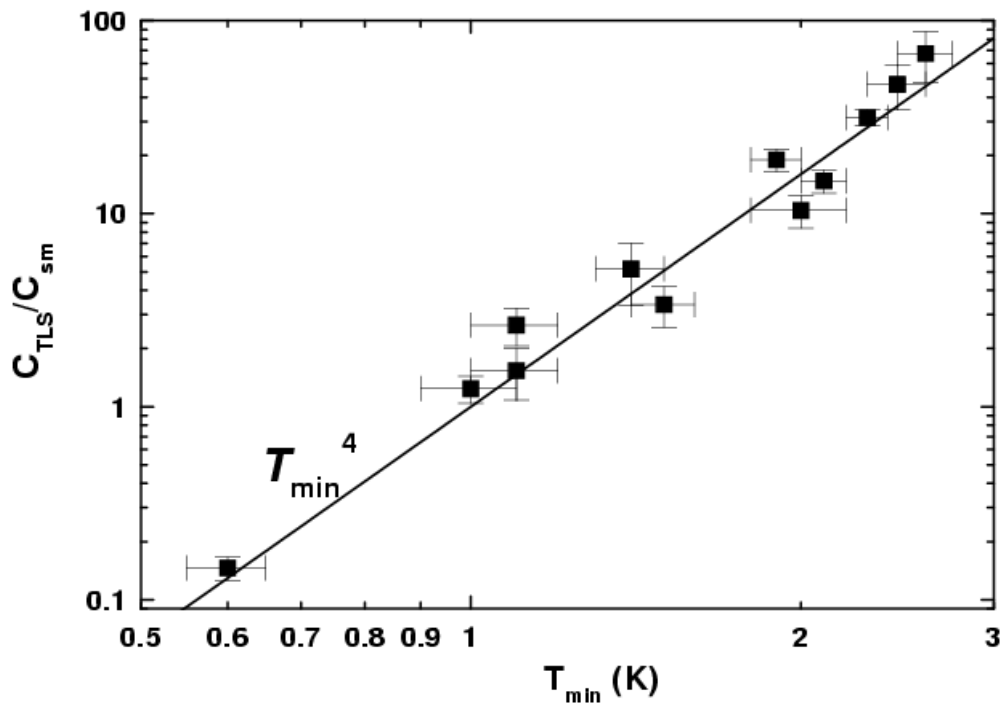

**Figure 4**. Check of the SPM: ratios $C_{TLS}/C_{sm}$ obtained from all the conducted fits, as a function of the observed $T_{min}$, in a log-log scale. The SPM prediction $C_{TLS}/C_{sm} \approx T_{min}^4$ is indicated by the solid line.

Secondly, another non-trivial assumption made has been that the linear contribution $C_1 \equiv C_{TLS}$ and the fifth-power contribution $C_5 \equiv C_{sm}$ are directly correlated as the SPM postulates, in contrast with other approaches. Dividing equation (4) by equation (5), the distribution constant $P_s$ cancels out, and one obtains $C_{TLS}/C_{sm} \approx T_{min}^4$, since evaluating the numerical factors, including the logarithmic factors, and some material-dependent constants, results in a numerical factor of order unity. Figure 4 displays the ratios $C_{TLS}/C_{sm}$ obtained from the fits shown in table 1, as a function of the observed $T_{\min}$. The ratio varies over three orders of magnitude, holding the approximate value $T_{\min}^4$, as indicated there by the solid line. Therefore, the assumed hypothesis based upon the SPM seems to be fully consistent.

In conclusion, the *quantitative* agreement demonstrated between *calorimetric* cubic coefficients and those predicted by the Debye theory from *elastic* data, once a consistent quadratic fit in the appropriate temperature range is carried out, strongly supports the applicability of the concept of acoustic phonons, and hence of the Debye theory, at least up to frequencies in the lower side of the boson peak. Furthermore, this finding also supports the fundamental idea of the Soft-Potential Model, namely, that quasilocal tunnelling states and soft vibrations can coexist with Debye acoustic modes.

ACKNOWLEDGEMENTS

I am very grateful to Professor Uli Buchenau for helpful discussions and for providing me with several specific-heat data files. This work has been supported by the Ministerio de Ciencia y Tecnología (Spain) within project BFM2000-0035-C02.